\documentclass{aastex631}

\usepackage{amsmath,amssymb,amsthm,amsfonts,bbold,gensymb}   
\usepackage{graphicx}   
\usepackage{pstricks}   
\usepackage{hyperref}   
\usepackage{listings}

\usepackage{mathrsfs}

\graphicspath{{./}{figures/}}

\shorttitle{PARSE: A Machine-Learning-Ready Active Region Dataset}
\shortauthors{Mathews and Thompson}

\begin{document}

\title{The Plasma-prescribed Active Region Static Extrapolation (PARSE) Dataset: A Machine-Learning-Ready Collection of Magnetohydrostatic Coronal Active Regions}

\author[0000-0002-8839-7860]{Nat H. Mathews}
\affiliation{NASA Goddard Space Flight Center, Greenbelt, MD, 20771, USA}
\correspondingauthor{Nat H. Mathews}
\email{n.h.mathews@nasa.gov}

\author[0000-0001-6952-7343]{Barbara J. Thompson}
\affiliation{NASA Goddard Space Flight Center, Greenbelt, MD, 20771, USA}

\begin{abstract}
    As Physics-Informed Neural Networks and other methods for full-vector-field construction or analysis become more prominent, a need has developed for a large set of simulated active regions for training, validation and testing purposes. We use a state-of-the-art magnetohydrostatic extrapolation method to develop a public dataset of over five thousand data cubes based on the Spaceweather HMI Active Region Patch (SHARP) library of active region magnetogram images. Each cube resolves the magnetic field vector and plasma forcing at approximately 100,000 scattered points that are adaptively clustered near the high-flux regions of the domain. This paper describes the methodology of construction of the Plasma-prescribed Active Region Static Extrapolation (PARSE) dataset, as well as its structure and how to access it.
\end{abstract}


\section{Introduction}

The Plasma-prescribed Active Region Static Extrapolation (PARSE) dataset consists of nearly-magnetohydrostatic (MHS), low-divergence extrapolations of photospheric boundary conditions. These boundary conditions are sourced from the Space weather HMI Active Region Patch (SHARP) image pipeline for the Helioseismic and Magnetic Imager (HMI) \citep{bobra2014}, and represent actual active regions which were present on the sun. Some of these active regions emitted flares or coronal mass ejections during their life cycle, and as such may encode interesting information in the volumetric magnetic vector field. By providing several possible magnetic configurations for each active region, we hope to facilitate a broad range of scientific pursuits. The varying structure and large quantity of the solutions may allow for the use of the dataset to train or validate physics-informed neural networks.

The extrapolation is performed by the Radial-Basis-Function Finite-Difference (RBF-FD) MHS solver \citep{mathews2022}. This allows for the resolution of the data on a scattered domain, which is leveraged to dynamically refine the solution. This allows for relatively high resolution near complicated structures, while keeping the overall memory requirements of the dataset small. For analysis which requires a regular lattice, the data can be easily interpolated by the user.

\section{Construction of the Dataset}

\subsection{Active Region Selection}

The SHARP dataset labels each active region as it rotates onto the face of the solar disk wtih a SHARP number. These active regions are then imaged with a six-minute cadence at high resolution. However, many statistical or machine-learning-based use cases of the PARSE dataset will require the samples to be independent of each other, and the SHARP images of a particular active region are closely correlated in time. To guarantee robustness against this temporal correlation in the dataset, we restrict ourselves to only one time frame from each SHARP number. This may result in more than one of a given active region, since they are re-numbered if they survive rotation across the far side of the sun, but we anticipate such a long cadence in potential repetitions to remediate any potential temporal correlation in the active region.

For each active region, we first discount any time frames whose flux-weighted centers are more than $60\degree$ (Stonyhurst) latitude or longitude away from disk center. Of the remaining frames (if any), the upper quartile are considered by total unsigned flux. Finally, the timeframe for modeling is chosen to be the one of these which is closest to disk center. In this way, we hope to obtain the best possible representative from each active region element to image and extrapolate.

\subsection{The Model Coordinate System}\label{sharpCoordinate}
The numerical model uses a cartesian coordinate system wherein $\hat{z}$ is the vertical direction ($z=0$ is defined as the photosphere), and $\hat{x}$ and $\hat{y}$ are the transverse components. We use the Disambiguated Lambert Cylindrical Equal-Area Projection vector field, which gives $B_r$, $B_\theta$ and $B_\phi$, which is provided by the SHARP repository directly. This disambiguation can include errors and assumptions, but for the purposes of this dataset they will be taken as given to allow easier comparison for the user with the original dataset.

The coordinate system in the SHARP is mirrored from the one in the model. To accommodate this, first the SHARP image is flipped vertically. Then $B_r$ is mapped directly to $B_z$, $B_\theta$ is mapped to $-B_x$ and $B_\phi$ to $-B_y$.

\subsection{Computational Node Layout}

A great power of the forward model we leverage for the extrapolation is its complete agnosticism with respect to computational node layout. It is not restrained to any kind of grid layout. We aim to take advantage of this property to cluster nodes near the spatially dynamic areas in the active region volume. To that end, we scatter nodes within the domain nearer high-flux regions of the boundary, while simultaneously scaling node density exponentially with height to cluster them near the lower boundary.

\begin{figure}
    \centering
    \includegraphics[width=0.4\textwidth]{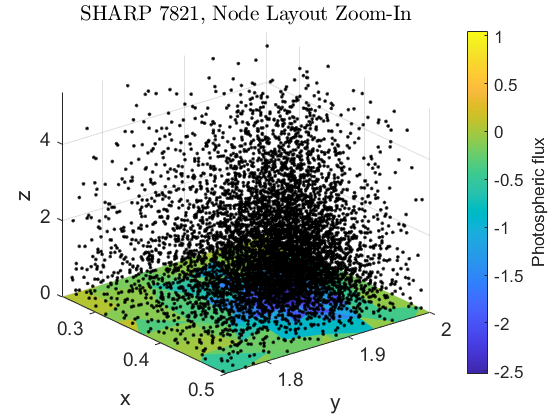}
    \caption{A zoom-in of a vertical column in the simulation of SHARP 7821, showing the node layout}
    \label{fig_nodes2D}
\end{figure}

A novel methodology to scatter nodes variably in space while retaining quasiuniform clustering has been the focus of recent research, and we apply state of the art advancing-front technique per \citet{vandersande2021}. A zoom-in on a portion of the domain of one simulation is given in Figure~\ref{fig_nodes2D}.

The clustering density of the method depends on an exclusion radius; the higher the function value, the sparser the node placement. We scale this quantity according to
\begin{equation}\label{electropotential}
    R(x,y,z) = \left((1-\tilde{B}_z(x,y)) L \cdot 0.01 + 0.015 \right) e^{z}
\end{equation}
where $\tilde{B}_z$ is a smoothed version of $|B_z-\text{median}(B_z)|$ constructed via a maximal binning window and normalized to have a maximum value of $1$, and $L$ is the ratio of the length of the longer transverse side of the domain to the shorter one (the domain is computationally normalized for the shortest side to be length $1$). Note that this algorithm generates nodesets with variable total numbers of nodes; $100,000$ is an upper bound for this quantity which is occasionally obtained, but most members of the PARSE dataset fall slightly below that threshold.

\subsection{Numerical Model}

The active region is extrapolated as a solution to the magnetohydrostatic equations, namely
\begin{equation}\label{eq_mhs}
\begin{split}
    \left(\nabla\times\mathbf{B}\right)\times\mathbf{B} &= \nabla P + \rho g \hat{z}\\
    \nabla\cdot\mathbf{B} &= 0
\end{split}
\end{equation}
We consider this a heterogeneous forced equation in terms of the conservative plasma forcing field $\mathbf{F}:=\nabla P + \rho g \hat{z}$.

The numerical model used for the extrapolations in this repository is described in detail in \citet{mathews2022}, and the curious reader is directed to that work. Here we discuss in detail only the model setup and determination of tunable parameters.

\begin{itemize}
 \item The photospheric boundary is informed by the SHARP, as described in Section \ref{sharpCoordinate}. Furthermore, a radiative condition is enforced at the upper boundary, $\partial_zB_z+B_z=0$. And the side boundaries have wave-permissible bounary conditions, $\partial_{nn}B_n+B_n=0$, $\hat{n}$ the normal vector to the given boundary.
 \item The numerical model uses a fixed hyperviscocity parameter to remediate numerical noise in the solution; a value of $\gamma=10^{-2}$ has been selected for this purpose.
 \item The model resolves and computes a set of nonphysical ``ghost nodes'' below the solar surface. These nodes are necessary for the extrapolation technique, but their fields are not considered physical, and they have been omitted from the published data.
 \item Finally, we find that the auxiliary equation solution included in the original method to remove any small divergence in the field converged poorly on the scattered domain, and has been omitted from calculations. This means that the domain is not necessarily totally free of magnetic divergence, but we note in Section \ref{analysis} that the effect of this omission is negligible.
\end{itemize}

\subsection{Plasma prescription}
The numerical model allows for different selections of $\mathbf{F}$ to be chosen and so poll the different possible topologies of the magnetic field configurations which can be extrapolated from the photospheric observations. However, the choice of plasma is not well-determined from the observations available. Indeed, determination of the correct magnetic field configuration is likely to be partially the task of the machine learning algorithm based on available auxiliary information.

Instead, we select a suite of \textit{possible} plasma distributions, determined by 100 scattered volumetric collocation points. Such a determination is considered possible if it satisfies the nullspace equation
\begin{equation}\label{nullspace}
\nabla P \cdot \mathbf{B} = 0
\end{equation}
at the $z=0$ surface. This is accomplished by taking the QR factorization of the transpose of the linear system in \eqref{nullspace} and taking a random subset of the columns of $Q$ which correspond to diagonal elements of $R$ below a threshold ($\varepsilon=10^{-3}$ was determined suitable for this purpose). This has the benefit of selecting plasmas which are orthonormal in terms of their volumetric collocation points, allowing more representative sampling of the model output space.

For purposes in which the plasma can be discounted, or the variations in magnetic topology not sufficiently interesting, a non-linear force-free extrapolation is included for each SHARP, obtained with an otherwise identical algorithm (the pressure is just set uniformly to $0$).

\section{Dataset Analysis}\label{analysis}

\begin{figure}
    \centering
    \includegraphics[width=0.3\textwidth]{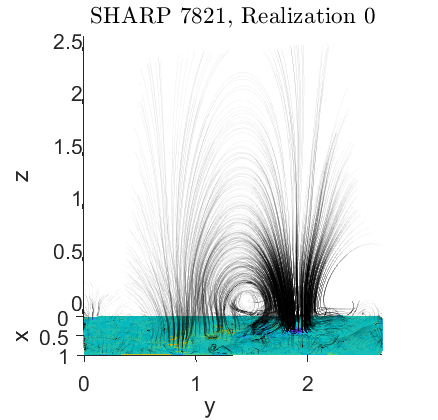}
    \includegraphics[width=0.3\textwidth]{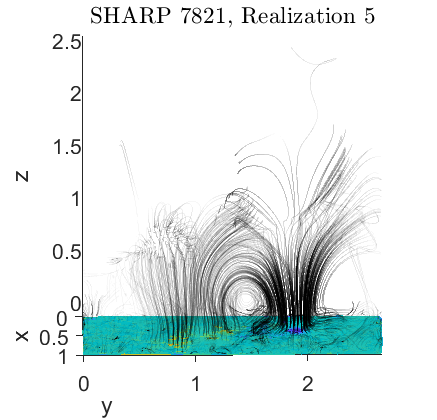}
    \includegraphics[width=0.3\textwidth]{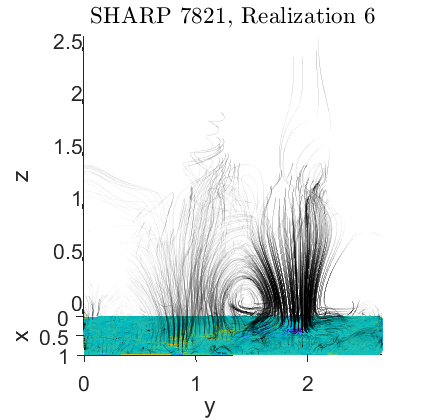}
    \caption{Three different extrapolations of SHARP 7821. The left is with no plasma prescription, and the middle and right are with different, orthogonal plasma pressures. The computational domain was evolved up to $z=5$, but has been cropped to $z=2.5$ for plotting. The photospheric plane has been colored according to magnetic flux.}
    \label{fig_extraps}
\end{figure}

We obtain strong convergence of MHS balance across all but a small subset of solutions. For most SHARPs, the solutions are topologically similar across plasma prescriptions, but with small deviations in loop height or volumetric structure. As can be observed in the three solutions in Figure~\ref{fig_extraps}, the forced fields are usually more complicated, pushed down closer to the photosphere, and have more transverse action, with field lines often leaving through the sides of the computational box. In some cases, such as the right panel in the same figure, the numerical hyperdiffusion was insufficient to remove all numerical artifacting, and we observe some small-scale field spirals. These are typically only in outbound open field lines, which are poorly constrained by the upper boundary conditions alone.

\section{Dataset Structure and Access}

Each solution is saved as a separate FITS file. The FITS file has a collection of Image Header Data Units (HDUs) which correspond to different physical parameters, described in Table~\ref{table_hdu}. Each is a one-dimensional array with entries corresponding to the scattered nodes. The header data of the primary HDU contains information about the simulation or active region, and is detailed in Table~\ref{table_header}.

\begin{table}
\begin{tabular}{| c | l |}
\hline
NAME & DESCRIPTION\\
\hline
MAIN & Primary HDU, containing no data but the header has\\ & meta-information about the SHARP or simulation.\\
\hline
BX & $x$ component of the magnetic field \\
\hline
BY & $y$ component of the magnetic field \\
\hline
BZ & $z$ component of the magnetic field \\
\hline
NODEX & $x$ coordinate at which the physical values are defined \\
\hline
NODEY & $y$ coordinate at which the physical values are defined \\
\hline
NODEZ & $z$ coordinate at which the physical values are defined \\
\hline
FX & $x$ component of the plasmatic forcing \\
\hline
FY & $y$ component of the plasmatic forcing \\
\hline
FZ & $z$ component of the plasmatic forcing \\
\hline
\end{tabular}
\caption{A table of the Header Data Units included in each FITS file.}\label{table_hdu}
\end{table}

\begin{table}
\begin{tabular}{| c | l |}
\hline
NAME & DESCRIPTION \\
\hline \hline
SIM\_N & The total number of nodes (i.e., the length of each column vector of physical parameters). \\
\hline
SIM\_L2 & Residual of the magnetohydrostatic simulation, in nodecount-normalized L2 norm.\\ & Provided as a proxy of simulation convergence; higher values may yield less physical solutions. \\ & Can be interpreted as a net force on the system exerted by the Lorentz and plasma forcing. \\ & The force-free field has a `NULL' entry.\\
\hline
LEN\_X & Size of the $x$ dimension of the computational box, in meters\\
\hline
LEN\_Y & Size of the $y$ dimension of the computational box, in meters\\
\hline
LEN\_Z & Size of the $z$ dimension of the computational box, in meters (the height)\\
\hline
LEN\_UNIT & Unit of LEN\_X, LEN\_Y and LEN\_Z\\
\hline
SHARPNUM & The SHARP number for the active region \\
\hline
ARNUM & The NOAA Active Region number for the active region, if it exists in the catalogue. \\ & This is a string, and sometimes is `MISSING' or contains more than one NOAA Active Region number. \\
\hline
TAI\_REC & The TAI date and time the active region was imaged, in the form `YYYY.MM.DD\_HH:MM:SS'. \\
\hline
USFLUX & Unsigned flux in the active region\\
\hline
AREA & The area of the de-projected SHARP patch in micro-hemispheres. \\
\hline
LON\_MIN & Minimum longitude of the active region (Stonyhurst). \\
\hline
LAT\_MIN & Minimum latitude of the active region (Stonyhurst). \\
\hline
LON\_MAX & Maximum longitude of the active region (Stonyhurst). \\
\hline
LAT\_MAX & Maximum latitude of the active region (Stonyhurst). \\
\hline
S\_NAXIS1 & Number of pixels along axis 1 of the original SHARP image\\
\hline
S\_NAXIS2 & Number of pixels along axis 2 of the original SHARP image\\
\hline
S\_CRPIX1 & X coordinate of disk center with respect to lower-left corner (in pixels) of the original SHARP image\\
\hline
S\_CRPIX2 & Y coordinate of disk center with respect to lower-left corner (in pixels) of the original SHARP image\\
\hline
S\_CRVAL1 & X origin of the original SHARP image: (0,0) at disk center\\
\hline
S\_CRVAL2 & Y origin of the original SHARP image: (0,0) at disk center\\
\hline
S\_CUNIT1 & unit of S\_CDELT1\\
\hline
S\_CUNIT2 & unit of S\_CDELT1\\
\hline
S\_CDELT1 & scale in the x direction of the original SHARP image\\
\hline
S\_CDELT2 & scale in the y direction of the original SHARP image\\
\hline
VERSION & A version number for the PARSE dataset, to allow distinguishing later updates to the simulations; \\ & the set described in this paper is Version 1.0.0.\\
\hline
\end{tabular}
\caption{A table of keywords associated with scalar-valued information about the active region, or the simulation. All but the first two are derived directly from the SHARP header.}\label{table_header}
\end{table}

Each SHARP is associated with a number of FITS files, indexed in the filename after the SHARP number. The $0$ index corresponds to the force-free solution, and the others after that are forced. For the 1.0.0 release, six forced solutions for each SHARP are provided.

The PARSE dataset is available open access on Zenodo (doi 10.5281/zenodo.8213061). The data generation code and example files that read the data are available on github at https://github.com/apt-get-nat/PARSE .

\section{Future Work}

The dataset is being continuously updated with more SHARPs. It should be possible to complete the dataset with an extrapolation of a single time frame from every extant SHARP active region by the end of 2024. We also wish to include a greater number of plasma-prescribed extrapolations for each observation to better capture the full range of possible topologies.

A future goal may also be to increase the resolution of the extrapolations. $100,000$ points can be considered a coarse resolution for some numerical use cases, and while it has proved sufficient for quantitative results \citep{mathews2020}, resolutions an order of magnitude higher would doubtless yield more physical magnetic fields.

\section{Acknowledgements}
This work was funded by the NASA Postdoctoral Program Fellowship, which is administered by Oak Ridge Association of Universities.

This work is based heavily on the SHARP dataset, and would not have been possible without the careful collection, stewardship and processing of that data by Monica Bobra and Stanford University.

Finally, this work was greatly aided by the temporary provision of computing resources by HP and Nvidia.

\bibliography{references}{}

\begin{thebibliography}{}
\expandafter\ifx\csname natexlab\endcsname\relax\def\natexlab#1{#1}\fi
\providecommand{\url}[1]{\href{#1}{#1}}
\providecommand{\dodoi}[1]{doi:~\href{http://doi.org/#1}{\nolinkurl{#1}}}
\providecommand{\doeprint}[1]{\href{http://ascl.net/#1}{\nolinkurl{http://ascl.net/#1}}}
\providecommand{\doarXiv}[1]{\href{https://arxiv.org/abs/#1}{\nolinkurl{https://arxiv.org/abs/#1}}}

\bibitem[{Bobra {et~al.}(2014)Bobra, Sun, Hoeksema, Turmon, Liu, Hayashi,
  Barnes, \& Leka}]{bobra2014}
Bobra, M.~G., Sun, X., Hoeksema, J.~T., {et~al.} 2014, Solar Physics, 289, 3549

\bibitem[{Mathews {et~al.}(2020)Mathews, Flyer, \& Gibson}]{mathews2020}
Mathews, N.~H., Flyer, N., \& Gibson, S.~E. 2020, The Astrophysical Journal,
  898, 70

\bibitem[{Mathews {et~al.}(2022)Mathews, Flyer, \& Gibson}]{mathews2022}
---. 2022, Journal of Computational Physics, 462, 111214,
  \dodoi{https://doi.org/10.1016/j.jcp.2022.111214}

\bibitem[{{van der Sande} \& Fornberg(2021)}]{vandersande2021}
{van der Sande}, K., \& Fornberg, B. 2021, SIAM Journal on Scientific
  Computing, 43, A242

\end{thebibliography}
\bibliographystyle{aasjournal}

\end{document}